\documentstyle[prl,aps,psfig,floats,twocolumn]{revtex} 
\begin{document} 
\draft 
\twocolumn[\hsize\textwidth\columnwidth\hsize\csname 
@twocolumnfalse\endcsname 
 
\newcommand {\be}{\begin{equation}} 
\newcommand {\ee}{\end{equation}} 
\newcommand {\bdm} {\begin{displaymath}} 
\newcommand {\edm} {\end{displaymath}} 
\newcommand {\bea} {\begin{eqnarray}} 
\newcommand {\eea} {\end{eqnarray}} 
\newcommand {\ra} {\rightarrow} 
\newcommand {\oo} {\infty} 
\newcommand{\bb} {\textbf}  
\newcommand{\e}{\emph}   
\newcommand{\lqu}{\left[} 
\newcommand{\rqu}{\right]} 
\newcommand{\lt}{\left(} 
\newcommand{\rt}{\right)} 
\newcommand{\la}{\langle} 
\newcommand{\ran}{\rangle} 
\newcommand{\sss}{\scriptsize} 
\newcommand{\bi}{\bibitem} 
 
\title{ A perturbative approach to  the Bak and Sneppen Model} 
\author{G. Caldarelli$^{1}$, M. Felici$^{1,2}$, A. Gabrielli$^1$ 
and L. Pietronero$^{1}$} 
\address{$^1$ INFM - Unit\`a di Roma 1 "La Sapienza", P.le A. Moro 2, 
00185 - Roma, Italy} 
\address{$^2$ Lab. PMC, Ecole Polytechnique 91128 Palaiseau, France} 
 
\maketitle 
\begin{abstract} 
We study here the Bak and Sneppen model, a prototype model for  
the study of Self-Organized Criticality.  
In this model several species interact  and undergo extinction  
with a power law distribution of activity bursts.  
Species are defined through their ``fitness' whose distribution in the system is  
uniform above a certain universal threshold. 
Run time statistics is introduced for the analysis of the dynamics  
in order to explain the peculiar properties of the model.  
This approach based on conditional probability theory, takes into account the  
correlations due to memory effects.  
In this way, we may compute analytically the value of the fitness  
threshold with the desired precision.  
This represents a substantial improvement with respect to the  
traditional mean field approach. 
\end{abstract} 
\pacs{02.50.-r, 64.60.Ak, 04.20.Jb} 
] 
\narrowtext 
 
\section{Introduction} 
In the recent years, many models mimicking the scale free behavior  
exhibited by natural phenomena like  
river basins \cite{rinaldo,riv}, fracture dynamics\cite{frac1,frac2},  
earthquakes\cite{eart}, have been extensively studied.  
The main features of these models are the lack of spatial and  
temporal characteristic scales, the evolution through 
intermittent bursts of activity and the absence of the  
fine-tuning of some parameter to reach the critical state.  
To describe all this, the concept of Self-Organized Criticality  
(SOC)\cite{btw} has been introduced,  
and many works have been devoted to clarify its real nature\cite{book}.  
 
We focus here on the evolution model introduced by Bak and  
Sneppen (BS) in 1993 \cite{bs}. 
This model is the prototype of a wide class of SOC models,  
characterized by a deterministic dynamics in a medium with quenched disorder.  
The quenched noise is independent on time and represents the  
disordered environment where the system evolves.  
BS model is defined by a discrete set of $L$ species arranged  
on a one dimensional network. 
Each species $i$ is defined by a ``fitness'' $x_i$ given by  
a real number in the interval $(0,1)$. 
Time evolution is discretized and at any time step the species  
with the lowest fitness is removed from the set.  
At the same time also the two species neighbors are removed.  
This should model a food web where extinction of one species affects  
also the survival of predation and prey species. 
Three new species with randomly extracted new fitness enter then the system. 
Hereafter, we shall call this process an ``updating'' of the species fitnesses. 
After a transient period, the system reaches spontaneously a  
stationary state characterized by two main features: (i) the fitnesses  
are uniformly distributed between a threshold value $x_c$ and $1$;  
(ii) the dynamics evolves as a sequence of \emph{critical avalanches}\cite{rev},  
whose duration $s$ is power-law distributed\cite{nota}:  
$P(s)\sim s^{-\tau}$, where $\tau=1.07$ \cite{bsdata}.  
 
To study in an analytical way this kind of processes, a method based on  
conditional probability  
has been recently introduced. This method, called Run Time Statistics (RTS),  
provides a powerful tool to  
study how the system stores information on the disorder during its evolution.  
This method has been also  
applied to the class of models derived by the Invasion Percolation  
\cite{wilkinson,pre} to compute the asymptotic behaviour of the histogram equation.  
We apply here RTS to the BS model in order to find an analytical approach  
to solve the model.  
 
As regards the analytical results available for the BS, the only successful  
approach has been the  
mean field approximation. Scale relations and an equation  
describing the hierarchy of the avalanches  
\cite{derr}, allow to reduce the number of independent exponents to one.  
A kind of $\epsilon$-expansion 
was also introduced to compute the critical exponents by performing an  
expansion around their mean field value.   
This novel approach allows to compute in a perturbative way the value  
of the critical threshold  linking it to the avalanche exponenti $\tau$.  
In the following sections we are going to introduce the  
main features of the model, the basic computation of the run time statistics  
and the result of this approach with respect to the BS. 
A preliminary letter with some of the results has already been published in  
Ref.\cite{maddalena}, here instead we are going to fully develop the derivation  
of such results. 
 
\section{Critical dynamics} 
The dynamics of the BS model can be viewed as a branching  
process of branching ratio $L$, if  
$L$ is the number of species in the system. At every time step, one of the $L$ quenched numbers assigned to the sites is selected to be updated.  
The same is done for the two nearest neighbors.  
It is then possible to represent this  
process as a tree-like picture where every node represents  
a state from which the system can reach  
$L$ possible states at the subsequent time step.  
Therefore at time $t$ there are $L^t$ possible states that are reached  
through $L^t$ different paths.  
Because of the deterministic nature of the model,  
time evolution is determined by the realization of the disorder $\{x\}$.  
This means that if we know the initial set of values $x_i$, $i=1,\ldots, L$, and the three numbers extracted at each time-step  
then we also know {\em a priori} the evolution of the system.  
Otherwise, to give a description of the behavior of the generic  
system, we should consider a statistical average  
over the possible realizations of the quenched disorder.  
RTS  provides an iterative algorithm to assign to each path  
(sequence of events) its statistical weight according to the laws of  
conditional probability. 
 
We consider the system in the critical state.  
In this stationary critical state of BS model  
almost all the fitnesses lie above a threshold $x_c$.  
The distribution function of the  
fitnesses in the critical state, called the {\em histogram}  
$\Phi (x)$, is found to be uniform above $x_c$, and almost  
zero below $x_c$. The critical state is characterized by a  
power law distribution of the duration $s$ of  
critical avalanches: $P(s)\sim s^{-\tau}$. 
Critical avalanches are defined as a sequence of events  
$x_{min}(s)<x_c$, where $x_{min}(s)$  
is {\em signal}, i.e. the value of the minimal fitness at time $s$ \cite{rev}. 
Then a critical avalanche begins each time the signal reaches a  
value larger than $x_c$. Because of the shape of the histogram, that is of  
order $\frac 1 L$ for $x<x_c$ and constant for $x>x_c$.  
In the critical state the signal reaches at most values near to $x_c$  
and the dynamics is a sequence of critical avalanches \cite{rev}.  
Then the site giving rise to a new avalanche, called the {\em initiator},  
has a fitness near to $x_c$.This means that when the initiator is selected, all  
the other sites have fitnesses larger than $x_c$. 
 
Since the critical avalanches are independent each other \cite{rev}, we  
can consider the dynamics within one generic critical avalanche. This is then  
a fair description of the dynamics of the system in the critical state.  
Let us define the set of active sites $A_t$ as the sites covered by the  
avalanche, i.e. the sites whose fitness has been updated at least once  
since the beginning of the avalanche. 
If we consider the dynamics within an avalanche,  
the possible events at time $t$ regard only the sites in $A_t$ because the  
selection of any other site would imply the end of the avalanche.  
Indeed, sites not belonging to 
$A_t$ have a fitness larger than $x_c$ (by definition of $A_t$  
their fitness remained unchanged since the first step of the avalanche).  
Therefore the evolution of the avalanche can be seen as a branching process  
where the branching ratio is not fixed.

\begin{figure}  
\centerline{\psfig{file=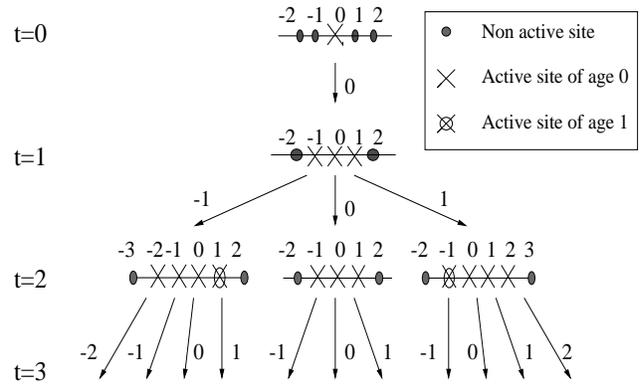,height=5cm,width=9cm,angle=0}}  
\caption{Diagrammatic plot of the first three steps  
in an avalanche tree. The initiator $i=0$ is selected at $t=0$.  
At each step one has to consider all the possible offsprings.  
Non-active sites are represented by a filled circle; updated active sites are  
represented by with a simple cross.  
Crossed empty circles represents instead active sites not updated at the previous time steps.}  
\label{tree}  
\end{figure}  
 
Without loss of generality, we take the time origin at the beginning of a critical avalanche, and the origin of the coordinates in the initiator site.  
At $t=0$ the initiator is the site with the smallest fitness (i.e. 
extremal rule) and its quenched variable (in the stationary state) has a value near to $x_c$.  
Since we are assuming that the stationary state in the system is the critical one, all the quenched numbers  
in the system are distributed following the stationary distribution $\Phi (x)$,  
apart from corrections of order $\frac 1 L$.  
Updating at time $t=0$ affects the initiator and the two nearest neighbors.  
$A_{t=1}$ is then composed by the three sites $\{-1,0,1\}$.  
The three variables corresponding to these sites are  
distributed following the uniform probability density $f_0(x)$.  
If the avalanche proceeds, there are three possible  
events leading to three different configurations for the system (see Fig.\ref{tree}).  
Then a path of length $2$ can be realized in the following three ways:  
i) growth of the initiator at time $t=0$ and then growth of its left neighbor at  
time $t=1$;  
ii) growth of the initiator at time $t=0$, and successive growth of the  
initiator at time $t=2$;  
iii) growth of the initiator at time $t=0$ and then growth of the right neighbor  
at time $t=1$.  
As the length $t$ (i.e. the number of steps) increases, the number $N_t$ of  
possible paths of that length increases fastly.  
For example, avalanches lasting two time steps can occur in the  
previous three ways, but there are 
eleven ways to form an avalanche of three time steps (see Fig.\ref{tree}). 
In table \ref{tabella} we report the number $N_t$ of possible paths whose length  
is $t$. It can be shown that $N_t$ grows roughly as $t!$. 
Given this picture of the stationary state, it is evident that a description of  
the model evolution can be achieved through a description of the growth paths.  
Run Time Statistics (whose approach is described in the next section) helps in  
sorting out the most probable paths, thereby extracting the available  
information on the process.   
\begin{table} 
\begin{center} 
\begin{tabular}{ |c|c|c|c|c|c|c|c|c|c|} \hline 
\textbf{t}&1&2&3&4&5&6&7&8&9\\ \hline 
$\mathbf{N_{t}}$&1&3&11&47&227&1215&7107&44959&305091\\ \hline 
\end{tabular} 
\caption{\small Number of possible paths as a function of time.} 
\label{tabella} 
\end{center} 
\end{table} 
 
\section{The RTS approach} 
Through the RTS we are able to compute the statistical weight of all the  
possible paths corresponding to a critical avalanche of a fixed  
duration\cite{luciano,matteo}. The RTS has been developed  
in order to extract the maximal information available from the knowledge of the  
dynamical history followed by the process.  
The information is stored in {\em effective} probability density functions (PDF's) for the variables $\{x\}$.  
These effective density functions at a certain time $t$, are used to compute the  
conditional probabilities of all the possible events at time $t+1$  
given the state at time $t$. The probability of a sequence of events  
(that is a fixed path), is then factorized in the product of these  
one-step-probabilities. The time  
dependent PDF's are obtained by applying the laws of  
conditional probability.  
At the beginning the only information available on the disorder  
is the probability density $f_0(x)$ from which the quenched variables are  
extracted. The conditional probability laws in the following steps modify the  
shape of the probability distribution once the previous history is known.  
 
In the case of BS, at every time step the smallest number is removed from the system.  
It is then intuitive that if the lifetime of a species, i.e. the number of tests  
the species survived in the search of the minimum, is large, its fitness is (probably) also large. 
Henceforth, the longer the lifetime the higher the probability to deal with a large value of the  
fitness $x$, and the smaller the probability to be removed.  
This concept can be easily formalized by using the theory of conditional probability.  
 
Let us introduce the time dependent PDF's $f_{i,t}(x)$ giving the probability  
density of the quenched variable of site $i$ at time $t$.  
If $m_{i,t}(x)$ is the probability density function of fitness at time $t$  
(assuming it corresponds to site $i$), we have that 
$m_{i,t}(x)dx$ is the probability that the fitness of site $i$ has a value between $x$ and  
$x+dx$  
(given that at time $t$ of a fixed path, $i$ is the minimum with fitness smaller than $x_c$; this last condition ensures that the system is still under the same critical avalanche).  
Then $m_{i,t}(x)$ is given by: 
\[m_{i,t}(x)\!= \!P\!\left(x<x_i<x+dx | x_i=min\{x\}_{A_{t-1}}, x_i<x_c\right)= 
\] 
\[=\frac{ P\left(x<x_i<x+dx \cap x_i=min\{x\}_{A_{t-1}}, x_i<x_c\right)}{P\left(  
x_i=min\{x\}_{A_{t-1}}, x_i<x_c \right)}=\] 
\begin{eqnarray} 
\label{emme} 
&\frac{1}{\mu_{i,t-1}}f_{i,t-1}(x)\prod_{k\in  
A_{t-1}-\{i\}}\int_{x}^{1} dx_{k}f_{k,t-1}(x_{k})\,  
&x\leq x_{c} \nonumber \\ 
& & x>x_{c}  
\end{eqnarray} 
 
Where we have defined the one step probability $\mu_{i,t}$: 
\[ 
\mu_{i,t}=P\left( x_i=min\{x\}_{A_t}, x_i<x_c \right)=\] 
\begin{equation}\label{mu} 
=\int_0^{x_c}dx f_{i,t-1}(x)\prod_{j\in A_{t-1}-\{i\}}\int_x^1 dx_j f_{j,t-1}(x_j).  
\end{equation} 
$\mu_{i,t}$ represents the probability that site $i$ has the minimal fitness at time $t$  
(smaller than $x_c$), given the path followed  
up to time $t-1$ ( all the information about the past steps is included in the effective  
PDF's $f_{j,t}(x)$).  
In both eq.(\ref{emme}) and eq.(\ref{mu}) we consider only the fitnesses 
$\{x\}_{A_t}$ of the active sites $A_t$ because the others actually do not participate in  
the dynamics, being larger than $x_c$. 
It is important to notice that 
\begin{equation} 
\sum_{i\in A_t}\mu_{i,t}<1 
\end{equation} 
This is because of the condition that the minimum fitness is less than $x_c$.  
The complementary probability  
$\left(1-\sum_{i\in A_t}\mu_{i,t}\right)$ is then the probability that the minimum is larger  
than $x_c$, that is that the avalanche stops at time $t$.

In a similar way we can be obtain the probability densities $f_{k,t}(x)$, with $k\in A_{t-1}$: 
 
\[f_{k,t}(x)\!=\!P\!\left( x<x_k<x+dx | x_i=min\{x\}_{A_{t-1}}, x_i<x_c \right)\!= 
\] 
\[=\frac{ P\left( x<x_k<x+dx \cap x_i=min\{x\}_{A_{t-1}}, x_i<x_c\right)}{P\left(  
x_i=min\{x\}_{A_{t-1}}, x_i<x_c \right)}= 
\] 
\begin{equation}\label{formule} 
=\left\{ 
\begin{array}{lll} 
\frac{1}{\mu_{i,t}}f_{k,t}(x)\int_{0}^{x}f_{i,t}(x_{i})dx_{i} \prod_{j}   
\int_{x_{i}}^{1}dx_{j}f_{j,t}(x_{j})\, x\leq x_{c}\\ 
\,\\ 
\frac{1}{\mu_{i,t}}f_{k,t}(x_{c})\int_{0}^{x_{c}}f_{i,t}(x_{i})dx_{i}\prod_{j}     
\int_{x_{i}}^{1}dx_{j}f_{j,t}(x_{j})\, x>x_{c}\\ 
\end{array} 
\right . 
\end{equation} 
where $j\in A_{t-1}-\{i,k\}$.  
In this way the effective PDF's are  
conditioned to the whole history from time $0$ (beginning of  
the avalanche) to time $t$ because  
of the step-by-step algorithm through which they are obtained.  
We also notice that if the minimum fitness is less  
than $x_c$, this implies that $f_{i,t}(x)=f_{i,t}(x_c)$ for $x>x_c$.  
As it has been pointed out in \cite{rev}, the dynamics involves only  
the quenched numbers below $x_c$, this is the reason why the system  
does not acquire information on the variables in the region $x>x_c$.

These formulae for the effective probability densities hold if we assume  
that the probability density of the whole set of variables $F_t(\{x\})$  
could be factorized in the product of the one-variable probability densities  
at any time. Actually this is true only at time $t=0$, when  
the $\{x\}$ are uncorrelated, while at later times the extremal dynamics  
induces correlations among them. Nevertheless the approximation 
\begin{equation} 
F_t(\{x\})\simeq \prod_j f_{j,t}(x) 
\end{equation} 
has proven to lead to results in good agreement with data from simulations \cite{euro1}, and to  
give rise to an error which is negligible for large  values of the system size $L$\cite{gab}. 
 
Once the one-step-probabilities $\mu_{i,t}$ have been computed, the probability of a given  
path $C_t$ (i.e. a fixed sequence of events from time 0 to time t) is given by:  
\begin{equation}\label{fac} 
W(C_t)=\prod_{t'=1}^{t} \mu_{i,t'} 
\end{equation} 
This probability is the probability of the path $C_t$ averaged on the  
disorder. 
 
Let us consider what would be the rigorous computation of this quantity. 
The extremal rule can be formulated by defining a growth probability $\eta_i(\{x\})$ 
for site $i$ to be selected, given by: 
\begin{equation} 
\eta_i(\{x\})=\prod_j \theta (x_j -x_i )=\Bigg\{  
{1 \quad x_{i}=min\{x\}\atop 0 \quad x_{i}\ne min\{x\} } 
\end{equation} 
 
\begin{figure}[t] 
\begin{center} 
\centerline{\psfig{file=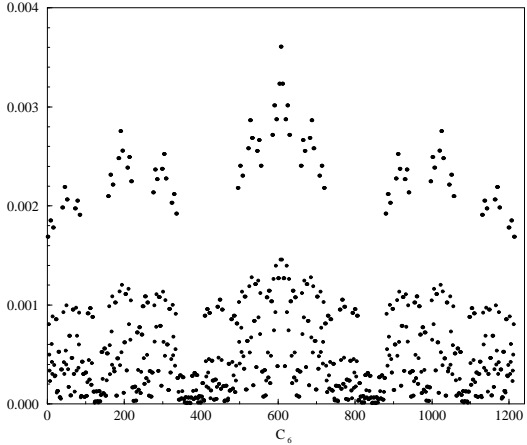,height=6cm,angle=270}} 
\caption{\small Statistical weights $W(C_6)$ of the paths $C_6$ of length 6.  
Paths are numbered from 1 to $N_6$ following the order in a tree-like diagram  
like the one shown in Fig.\ref{tree}. } 
\label{probcamm} 
\end{center} 
\end{figure}

\begin{figure}[t] 
\begin{center} 
\centerline{\psfig{file=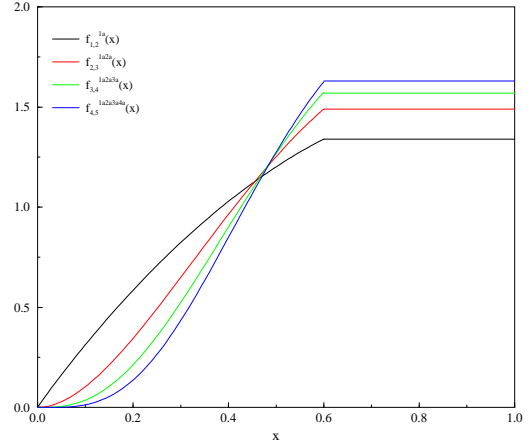,height=6cm,angle=270}} 
\caption{\small Effective probability density of the site $1$ at  
different times of the path \textbf{1a2a3a4a}.  
The probability densities are labeled with the  
indexes $\tau ,t$ giving the ``age'' of the quenched number,  
and the time step of the path.} 
\label{rtstime} 
\end{center} 
\end{figure}

Then the probability of a given path $C_t$ is given by: 
\begin{equation} 
\Omega_{C_t}(\{x\}) =\prod_{t'=1}^{T}\eta_{i_{t'}}(\{x\}_{A_{t'}}) 
\end{equation} 
which can assume only the values 1 or 0. The exact computation of the \emph{mean} 
probability of the path $C_t$ consists in taking the average over the realizations 
of the disorder. This average cannot be factorized: this means that it is not possible  
to obtain the mean probability of a path $C_t$ by  
simply multiplying the mean probability of the path $C_{t-1}$, corresponding to the first $t-1$ 
steps of path $C_t$, for the mean  
probability of the last step. 
On the contrary, the weight $W(C_t)$ is factorized as in eq.(\ref{fac})  
because the one-step probabilities $\mu_{i,t}$ are \emph{conditional}  
probabilities.  
The weights $W(C_t)$ of the paths of length $6$ are plotted in  
Fig.(\ref{probcamm}). In Fig(\ref{rtstime}) are plotted the effective  
probability densities $f_{i,t}(x)$ of a site $i$ for different values of $t$:  
the site chosen is not updated from time $t=1$ to time $t=4$,  
the corresponding PDP is then modified in such a way to approach the  
histogram shape.

\section{The histogram equation} 
We now introduce an equation for the histogram $\Phi_t(x)$,  
that represents the average probability  
density of the quenched variables at time $t$.  
In the limit $t\rightarrow \infty$ we obtain an equation for the  
stationary histogram.  
Let us introduce the function $h_t(x)$ defined in the following way: 
\begin{equation} 
h_t(x)=L\Phi_t(x) 
\end{equation} 
where $L$ is the size of the system.  
Then $h_t(x)dx$ is the average number of quenched variables in the system in  
the interval $[x,x+dx]$ at time $t$. Since the species updated at each  
time step are the minimal one with the two nearest neighbors,  
we can write a balance equation : 
\begin{equation} 
h_{t+1}(x)=h_t(x)-m_{t+1}(x)-\left( f_1(x)+f_2(x)\right) +3f_0(x) 
\end{equation} 
where $m_t(x)$ is the probability density of the minimal variable;  
$f_{1,2}(x)$ are the probability densities that the two nearest  
neighbors variables would have at time $t+1$ if they were not updated.  
The PDF's $f_{1,2}(x)$ are given by the first line of  
eq.(\ref{formule}) with $k=i-1$ and $k=i+1$ respectively. 
Because of the self averaging property of this function the result obtained by  
taking the asymptotic limit coincides with the one obtained by averaging over  
the possible realizations of the disorder: 
\begin{equation} 
\lim_{t\rightarrow\infty}h_t(x)=L\Phi(x)=L\langle 
\Phi(x)\rangle_{\{x\}} 
\end{equation}  
We then obtain: 
\begin{equation}\label{equaisto} 
\langle m(x)\rangle +\langle f_1(x)+f_2(x)\rangle -3=0 
\end{equation} 
where $f_0(x)=1$. 
 
To compute the average $\langle f_1(x)+f_2(x)\rangle$ we use the weights $W(C_t)$  
obtained by applying the RTS algorithm: $\langle f_1(x)+f_2(x)\rangle$ is given by  
averaging over the paths $C_t$, weighed with the $W(C_t)$ 
\begin{equation}\label{effemedie} 
\langle f_1(x)+f_2(x)\rangle= \frac {\sum_{t=1}^{\infty}\sum_{C_{t}}W(C_{t}) \left[ 
f_{1}^{C_{t}}(x)+ f_{2}^{C_{t}}(x)\right]}{\sum_{t=1}^{\infty}\sum_{C_{t}}W(C_{t})} 
\end{equation} 
To compute this quantity to the order $n$, we perform an exact enumeration  
of the paths of length $t\leq n$. For each path $C_t$, $f_{1,2}^{C_t}(x)$ are  
computed by iterating the formulae in eq.(\ref{formule}) to obtain the effective  
probability densities at time $t+1$ of the nearest  
neighbors sites of the site selected to grow at time $t$ of the path $C_t$.  
The sum in eq.(\ref{equaisto}) contains also terms proportional to $\Phi(x)$.  
This happens when one of the nearest neighbors of  
the extremal site does not belong to the set of active sites,  
its probability density is $\Phi(x)$. Moreover  
$x_c$ appears explicitly in eq.(\ref{effemedie}i) because both the probability  
densities $f_{1,2}(x)$ and the weights $W(C_t)$ depend on it,  
(since $x_c$ has been introduced as a parameter by  
imposing that the system is under an $x_c$-avalanche).  
 
The minimum probability density is averaged in a different way, following a mean field argument.  
We now consider the Generalized Run Time Statistics \cite{gab}, which gives the  
correct form for the effective probability densities in the case of a  
stochastic dynamics.  
Let us suppose to have a growth probability $\eta (\{x\})$ depending on the  
quenched disorder different from extremal rule.  
Also in this case the system stores information during the evolution.  
The eq.{\ref{formule}) must now be modified in order to take into  account  
the probability $\eta_i(\{x\})$. The one-step probability is obtained in the  
following way  
\begin{equation} 
\mu_{i,t}\!=\!\int_{0}^{1}\!f_{i,t}(x_{i})dx_{i} \int_{0}^{1}\!\!\cdots\!\!\int_{0}^{1}\! 
\prod_{j\in A_{t}-\{i\}}\! dx_{j}f_{j,t}(x_{j})\eta_{i}(\{x\}_{A_{t}}) 
\end{equation} 
where now $A_t$ is given by all the sites in the system.  
The probability density of the minimum site is consequently given by: 
\[ 
 m_{i,t+1}(x)=\frac{1}{\mu_{i,t}}\int_{0}^{1}dx_{i}\delta(x_{i}-x)f_{i,t}(x_i) \cdot \] 
 \begin{equation} \label{emme2} 
\qquad \cdot \int_{0}^{1}\cdots\int_{0}^{1}\prod_{k} \eta_{i}(\{x\}_{A_{t}})f_{k,t}(x_{k})dx_k 
 \end{equation} 
where $i_t$ is the quenched number of the site selected at time $t$. 
 
We now consider the growth probability given by: 
\begin{equation}\label{eta} 
\eta_{i}(\{x\}_{A_{t}})=\frac{e^{-\frac{x_{i}}{T}}}{\sum_{j\in  
A_{t}}e^{-\frac{x_{j}}{T}}} 
\end{equation} 
In this way in the limit $T\rightarrow 0$ we recover the extremal rule: 
\begin{equation} 
\lim_{T\rightarrow 0}\frac{e^{-\frac{x_{i}}{T}}}{\sum_{j=1}^{L}e^{-\frac{x_{j}}{T}}}=\Bigg\{  
{1 \quad x_{i}=min\{x\}\atop 0 \quad x_{i}\ne min\{x\} } 
\end{equation} 
We then substitute the expression eq.(\ref{eta}) in eq.(\ref{emme2}) to  realize  
the average over the disorder and then the limit $T\rightarrow 0$.  
 
We average on the paths $C_t$ of length $t$ and then consider the limit $t\rightarrow  
\infty$. The average on the paths $C_t$ is taken by averaging first on the last step and then on the paths $C_{t-1}$: 
\begin{equation} 
\langle m_{i,t}(x)\rangle_{C_t}=\langle \langle m_{i,t}(x)\rangle_{i_{t}}\rangle_{C_{t-1}} 
\end{equation} 
The first average gives: 
\[ 
\langle  m_{i,t}(x)\rangle_{i_{t-1}}=\sum_{i=1}^{L}\mu_{i,t} \Bigg( \frac{1}{\mu_{i,t}} f_{i,t}(x)  
\cdot \] 
\[ 
\qquad \cdot  \int_{0}^{1}\cdots\int_{0}^{1}\prod_{j\neq i} dx_j f_{j,t}(x) \frac{  
e^{-\frac{x}{T} }  }  { e^{-\frac x T } +\sum_{k\neq i} e^{-\frac{x_k} T  } } \Bigg)\] 
Then, averaging over the paths $C_t-1$ we obtain: 
\[ 
\langle  \langle  m_{i,t}(x)\rangle_{i_{t}}\rangle_{C_{t-1}}=\] 
\[= \sum_{i=1}^{L} \langle f_{i,t}(x) \int_{0}^{1}\cdots\int_{0}^{1} \prod_{j\neq i}dx_j  
f_{j,t}(x)  
 \frac{ e^{-\frac{x}{T} }  }  { e^{-\frac x T } +\sum_{k\neq i} e^{-\frac{x_k} T  } }\rangle  
\approx \] 
\begin{equation} 
\approx \sum_{i=1}^{L}\langle f_{i,t}(x)\rangle  
   \int_{0}^{1}\cdots\int_{0}^{1} \prod_{j\neq i}dx_{j}\langle f_{j,t}(x)\rangle 
\frac{e^{-\frac x T}}{e^{-\frac x T}+\sum_{k\neq i}e^{-\frac {x_{k}} {T}}} 
\end{equation} 
Taking the limit $t\rightarrow \infty$ we have $\langle f_{j,t}(x)\rangle= \Phi(x)$: 
 
\begin{equation}\label{ulti} 
\langle  m(x)\rangle =L\Phi(x)\int_{0}^{1}\cdots\int_{0}^{1}\left[\prod_{j\neq  
i}dx_{j}\Phi(x_{j})\right] \frac{e^{-\frac x T}}{e^{-\frac x T}+\sum_{k\neq i}e^{-\frac  
{x_{k}} {T}}} 
\end{equation} 
Let us consider the integral in the RHS of  eq.(\ref{ulti}): it is the average of the function  
$g(\{x_k\})=\frac{e^{-\frac x T}}{e^{-\frac x T}+\sum_{k\neq i}e^{-\frac  
{x_{k}} {T}}}$, which is a function of the $(L-1)$ variables $\{x_k\}_{k\neq i}$ 
. If we put $z_j=e^{-\frac {x_j}{T} }$ and $Z=\sum_k z_k$, according to the central limit   
theorem the deviation from the mean value $\langle Z\rangle = (L-1)\langle z \rangle$ is  
negligible. We can then make the approximation $\langle F(Z)\rangle\approx F(\langle  
Z\rangle )$: 
\[ 
\int_{0}^{1}\cdots\int_{0}^{1}\prod_{j\neq i}dx_{j}\Phi(x_{j})\frac{e^{-\frac x  
T}}{e^{-\frac x T}+\sum_{k\neq i}e^{-\frac {x_{k}} {T}}}\approx \] 
\[ 
\approx \frac{e^{-\frac x  
T}}{e^{-\frac x T}+(L-1)\langle e^{-\frac {x_{k}} {T}}\rangle} 
\] 
We now introduce the parameter $x_c$ by defining $\langle z\rangle =\langle e^{-\frac {x}  
{T}}\rangle =e^{-\frac {x_c} {T}}$. We finally obtain: 
\begin{equation}\label{emmefinale} 
\langle m(x)\rangle =L\Phi (x)\frac {1}{1+(L-1)e^{\frac {-(x_{c}-x)}{T}}} 
\end{equation} 
We can see that if we assume for $\Phi(x)$ the following behavior: 
\begin{eqnarray} 
\Phi(x)&=&\Bigg\{ {{O(\frac 1 L) \qquad x\leq x_{c}}\atop{O(1)\qquad x>x_{c}}  } 
\end{eqnarray} 
then the average of the minimum density function given above has the expected behavior, in the limit $T\rightarrow 0$: 
\begin{eqnarray} 
\langle m(x)\rangle &=&\Bigg\{ {{O(1)\qquad x\leq x_{c}}\atop{0\qquad\qquad x>x_{c}}} 
\end{eqnarray} 
Thus the parameter $x_c$ introduced above is the critical threshold  
in the histogram function. 
 
We can now turn to the eq.(\ref{equaisto}).  
After use of eq.(\ref{emmefinale}) and eq.(\ref{effemedie}),  
the equation takes the following form: 
\[L\Phi (x)\frac {1}{1+(L-1)e^{\frac {-(x_{c}-x)}{T}}}+\] 
\begin{equation}\label{equaisto2} 
\qquad +\frac{A^{(n)}(x_c)\Phi(x)+B^{(n)}(x,x_c)}{D^{(n)}(x_c)}-3=0 
\end{equation} 
where the coefficients $A^{(n)}(x_c)$, $B^{(n)}(x,x_c)$  
and $D^{(n)}(x_c)$  are given by   
eq.(\ref{effemedie}) with truncation at order $n$.  
There is only one value of $x_c$ for which $\Phi (x)$ is  
normalized. This value gives the value of the BS threshold. 
  
It is easy to verify that the first order of approximation,  
$n=1$, corresponds to the mean field approximation.  
Indeed, the first order of eq.(\ref{effemedie}) is: 
\begin{equation}\label{orduno} 
\langle f_1(x)+f_2(x)\rangle ^{(1)}= 2\Phi(x) 
\end{equation} 
This is essentially the random neighbors assumption   
since in this way all the correlations among the  
species in the avalanche are neglected. By substituting  
eq.(\ref{orduno}) in eq.(\ref{equaisto2}) we obtain: 
\begin{equation} 
\Phi (x)=\frac {3}{2+\frac{L}{1+(L-1)e^{-\frac{(x_{c}-x)}{T} }}} 
\end{equation} 
that in the limit $T\rightarrow 0$, becomes: 
\begin{eqnarray} 
\label{lala} 
\Phi(x)&=& \Bigg\{  { {\frac 3 L \qquad\qquad x\leq x_{c}}\atop{\frac 3 2  
\qquad\qquad x>x_{c}}} 
\end{eqnarray} 
Then, imposing the normalization condition 
\begin{equation}\label{norm} 
\int_{0}^{1}\Phi(x)dx=1  
\end{equation} 
 we obtain  
$\frac 3 2 (1-x_{c})=1$, verified for $x_c =\frac 1 3$, which is the value obtained  
in the mean field case \cite{derr}.  
It is worth to notice that in the eq.(\ref{lala}) we have obtained  
analytically the behavior $\Phi(x)\sim \frac 1 L$ under $x_c$. 
 
We solved numerically eq.(\ref{equaisto}) from order $n=2$ to order $n=7$.  
The values of $x_c$ obtained for different values of $n$ are plotted in  
Fig.\ref{fig2}. By considering part of the correlations among species  
$x_c$ becomes larger than the one obtained in the mean field  
approximation ( $x_c=1/3$ ). 
The best evaluation is $x_c (n=7 )\simeq 0.465$, larger than the mean  
field result but still quite far from the value obtained from simulations.  
Nevertheless, it is possible to  
verify that the behavior of $x_c (n)$ is compatible with an asymptotic value  
$x_c (n\rightarrow \infty )\simeq 0.66$.  
We made a fit with the fitting function $x_c (n)=0.66-ax^b$ 
which is found to be well compatible with the given  
asymptotic value (see Fig.\ref{fig2}). 
The fit values are $a=0.291\pm 0.003$ and $b=0.20\pm 0.03$.  
The small value of $b$ is due to the fact that the avalanche  
duration distribution $P(s)$ is characterized by a small exponent  
($\tau\simeq 1.07$), henceforth all the sizes $s$ are important for statistics. 
 
One can use $x_c (n\rightarrow \infty )$ to evaluate both the avalanche exponent $\tau$ and  
the average minimum distribution $m(x)\equiv \left< m_i(x)\right>$. 
The exponent $\tau$ can be obtained from the function $\Omega (t)=\sum_{C_t}W(C_t)$ for 
$t$ ranging from 1 to the maximal possible $n$. This function is proportional to the probability 
that the avalanche lasts at least $t$ time-steps. Thus, in the scaling regime  
$\Omega (t)\sim t^{-\tau+1}$. Making this hypothesis, and substituting the value 
 $x_c (n\rightarrow \infty )$ in the expressions giving the weights $W(C_t)$, one finds 
$\tau (n=7)\simeq 1.05$ (see Fig.\ref{fig2}), which is in agreement with the known numerical value. 
 
Finally, we can obtain an approximation of the probability density function of  
the minimal fitness $m(x)$ in the stationary state.  
Let us consider the eqs.\ref{emmefinale} and  
\ref{equaisto2}. If we take the limit $T\rightarrow 0$ in eq.\ref{emmefinale} 
we obtain: 
\begin{equation} 
\left< m(x)\right> =L\Phi (x)\theta (x_c -x) 
\end{equation} 
If we now turn to eq.(\ref{equaisto2}) and we solve it for $\Phi(x)$ using the  
value $x_c (n\rightarrow \infty )$, we obtain the function $m(x)$ reported in  
Fig.\ref{fig3}. 
In the same figure, this result is compared with the numerical distribution of the minimal fitnesses obtained in the numerical simulations.  
In spite of the strong approximation (the paths considered are only those of  
length $\leq 7$), the agreement is quite good.  
 
In conclusion, this paper presents a perturbative
approach to the BS
model, based on the probabilistic framework called Run Time Statistics.
The detailed derivation of the  
self-organized threshold $x_c$, the avalanche exponent $\tau$, and
the stationary distribution of minimal fitnesses $m(x)$ is presented here.
Through RTS we are able to improve the agreement between the numerical and
the theoretical values found for this model. 
 
\begin{figure}[t] 
\begin{center} 
\centerline{\psfig{file=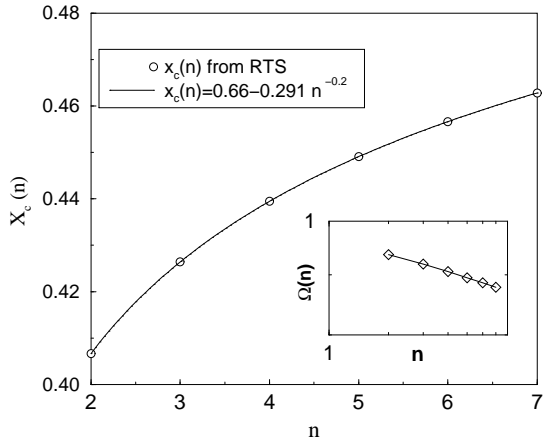,height=6cm}} 
\caption{\small Empty points represent the values of $x_c$  
obtained by the application of the RTS algorithm from $n=2$ to $n=7$. 
The continuous line represents the fit curve $x_c(n)=0.66-ax^b$ with 
$a=0.291\pm 0.003$ and $b=0.20\pm 0.03$. 
The insert shows the behavior of $\Omega(n)$ up to $n=7$. Assuming  
$\Omega(n)\sim n^{-\tau+1}$ 
as a good approximation also for small $n$, one finds $\tau\simeq 1.05$.} 
\label{fig2} 
\end{center} 
\end{figure}

\begin{figure} 
\centerline{\psfig{file=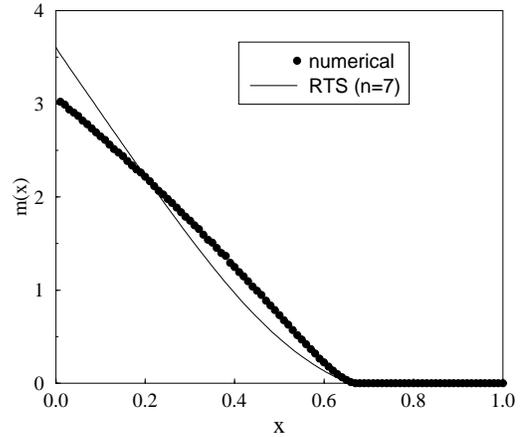,height=6cm,angle=-90}} 
\caption{The continuous line gives the stationary distribution $m(x)$ of  
the minimal fitnesses evaluated through the application of the RTS algorithm, 
considering all the possible avalanche paths up to a maximal time $n=7$  
after the selection of the initiator, and assuming  
$x_c(n\rightarrow\infty)=0.66$. 
The points gives the numerical behavior evaluated in extensive simulations.}  
\label{fig3}  
\end{figure}  



\begin{thebibliography}{99}
\bibitem{note}  
It is in principle possible to define a more general stochastic dynamics  
by assigning to each site of the lattice  
a probability to be selected for updating that depends on the disorder, that is on the set of  
numbers $\{x_i\}, i=1,\ldots,L$.  
The extremal rule corresponds to the particular case when this probability is one for the site  
with minimal fitness in the system, and zero for all the others. In such a case, the dynamics is  
deterministic. In general,the main feature of the models with quenched disorder is the presence of memory  
effects.
\end{thebibliography}
\end{document}